\documentclass[12pt,preprint]{aastex}

\def\gsim{ \lower .75ex \hbox{$\sim$} \llap{\raise .27ex \hbox{$>$}} }
\def\lsim{ \lower .75ex\hbox{$\sim$} \llap{\raise .27ex \hbox{$<$}} }

\begin{document}

\title{Explaining the High Energy Spectral Component in GRB 941017}

\author{Jonathan Granot\altaffilmark{1} and Dafne Guetta\altaffilmark{2}\altaffilmark{,}\altaffilmark{3}}

\altaffiltext{1}{Institute for Advanced Study, Olden Lane, Princeton, NJ 08540; granot@ias.edu}
\altaffiltext{2}{Osservatorio astrofisico di Arcetri, L.E. Fermi 2, Firenze, Italy; dafne@arcetri.astro.it}
\altaffiltext{3}{Racah Institute for Physics, The Hebrew University, Jerusalem 91904, Israel}
\begin{abstract}
  
  The gamma-ray burst (GRB) of October 17, 1994 (941017), showed a
  distinct high energy spectral component extending from $\lesssim\;$a
  few MeV to $\gtrsim 200\;$MeV, in addition to the typical GRB
  emission which peaked at $\lesssim\;$a few hundred keV. The high
  energy component carried at least $\sim 3$ times more energy than
  the lower energy component. It displayed an almost constant flux
  with a rather hard spectrum ($F_\nu\propto\nu^{-\alpha}$ with
  $\alpha\sim 0$) from $\lesssim 20\;$s into the burst up to $\sim
  200\;$s, while the duration of the GRB, where $90\%$ of the energy
  in the lower energy component was emitted, was only $77\;$s.  Such a
  high energy component was seen in only one out of $\sim 30$ GRBs in
  which a similar component could have been detected, and thus appears
  to be quite rare.  We examine possible explanations for this high
  energy spectral component and find that most models fail. The only
  emission region that provides the right temporal behavior is the
  reverse shock that goes into the GRB ejecta as it is decelerated by
  the ambient medium, or possibly the very early forward shock while
  the reverse shock is still going on. The best candidate for the
  emission mechanism is synchrotron self-Compton emission from the
  reverse shock. Even in this model the most natural spectral slope is
  only marginally consistent with the observed value, and some degree
  of fine tuning is required in order to improve the agreement. This
  might suggest that an additional or alternative emission mechanism
  is at work here. A prediction of this interpretation is that such a
  high energy component should be accompanied by a bright optical
  transient, similar to the one observed in GRB 990123.

\end{abstract}


\section{Introduction}
\label{sec:intro}

The spectrum of the prompt emission in gamma-ray bursts (GRBs) is
usually well described by the empirical Band function (Band at al.
1993) which features two power laws that join smoothly near the
typical photon energy $E_{\rm peak}$, where $\nu F_\nu$ peaks.  In the
vast majority of cases $E_{\rm peak}$ ranges between a few tens of keV
to a few MeV. This familiar and well studied spectral component is
most likely from synchrotron radiation of relativistic electrons in a
strong magnetic field, as is suggested by the recent measurement of a
very high degree of linear polarization in the prompt $\gamma$-ray
emission of GRB 021206 (Coburn \& Boggs 2003).

The highly variable light curve of most GRBs suggests that they
originate from internal shocks (Rees \& M\'esz\'aros 1994; Sari \&
Piran 1997) within a variable relativistic outflow from a compact
source.  When the ejecta sweeps up enough external medium, it is
decelerated by a reverse shock that propagates back into the ejecta,
while a strong relativistic forward shock is driven into the ambient
medium. The forward shock is believed to produce the afterglow
emission observed in the X-ray, optical and radio over a time scale of
days, weeks, and months, respectively, after the GRB (for a review see
van Paradijs, Kouveliotou \& Wijers 2000). The reverse shock produces
a much shorter lived emission, that rapidly decays after the shock
finishes crossing the shell of ejecta, which typically occurs on a
time scale similar to the duration of the GRB, $T_{\rm GRB}$.  The
synchrotron emission from the reverse shock is expected to peak around
the near UV or optical (Sari \& Piran 1999a).  A very bright optical
transient, that reached 9th magnitude in the optical, was observed
during the prompt $\gamma$-ray emission of GRB 990123 (Akerlof et al.
1999), and was successfully interpreted as emission from the reverse
shock (Sari \& Piran 1999b; M\'esz\'aros \& Rees 1999). Similar
optical observations within the first $\sim 100\;$s in other GRBs
produced only upper limits of $\sim 13-15\;$mag, (Kehoe et al. 2001).

In a recent paper, Gonz\'alez et al. (2003, hereafter G03) presented
new data for GRB 941017 which shows clear evidence for a distinct high
energy spectral component, in addition to the usual lower energy
spectral component. The latter is well fit by a band function with
$E_{\rm peak}$ decreasing from $\sim 500\;$keV to a few tens of keV
during the GRB, and is similar to other GRBs, suggesting a common
origin; it emitted $90\%$ of its energy over a time $T_{\rm
  GRB}=77\;$s. The high energy component appears $\sim 10-20\;$s after
the start of the GRB\footnote{There is a hint in the data of G03 that
  it may also be present from the very start.} and displays a roughly
constant flux with a relatively hard spectral slope
($F_\nu\propto\nu^{-\alpha}$ with $\alpha\sim 0$) up to $\sim 200\;$s.
The very different temporal behavior of the two components may suggest
a different physical origin.

Such a bright high energy component appears to be quite rare in GRBs.
EGRET observations of $>100\;$MeV photons from four other GRBs, as
well as 25 other GRBs that were bright at $300\;$keV and were also
detected by TASC 
(Total Absorption Shower Counter on board the Compton Observatory) 
showed a high energy emission that is consistent with the
single spectral component observed by BATSE (G03).  Therefore, any
model that tries to explain the high energy component in GRB 941017
should be able to explain at the same time why a similar component is
not seen in most GRBs.

In this {\it Letter} we analyze relevant physical mechanisms that
might produce such a high energy spectral component, and examine their
ability to explain this observation.  The possible explanations are
presented according to the relevant emission region, namely either the
internal shocks (\S \ref{sec:IS}), or the external shock (\S
\ref{sec:RS}) which includes the reverse shock and the early emission
from the forward shock. Different emission mechanisms are considered
for each region. Our conclusions are discussed in \S \ref{sec:dis}.

\section{Internal Shocks}
\label{sec:IS}

An important difficulty that arises when trying to explain the high
energy component as emission from the internal shocks is that in this
case it is attributed to the same shocks that emit the lower energy
component, and it is therefore expected to show a similar temporal
behavior.  However, in GRB 941017 the high energy component is almost
constant in time from $\lesssim 20\;$s to $\sim 200\;$s, while the
lower energy component decays on a shorter time scale, with $T_{\rm
  GRB}=77\;$s (G03). This poses a serious problem to most of the
emission mechanisms mentioned below.

Let us first examine synchrotron self-Compton (SSC) emission, i.e. the
inverse-Compton (IC) up-scattering of synchrotron photons by the same
electrons that emit the synchrotron radiation (the latter is
identified here with the lower energy spectral component). The SSC
spectrum is similar to the synchrotron spectrum, with the peak of $\nu
F_\nu$ at a frequency and a flux higher by a factor of $\sim\gamma_e^2
\sim 10^5$ and $Y$, respectively, where $Y$ is the Compton
y-parameter.  While this might reasonably account for the spectral
slope of the high energy component, the peak energy is around $\sim
10-100\;$GeV, implying $Y\gtrsim 10^3$, and $\sim 3$ orders of
magnitude more energy in the high energy component, compared to the
lower energy component, which is among the brightest BATSE
bursts.\footnote{If this burst was very close, i.e. at a redshift
  $z\ll 1$ instead of the typical $z\sim 1$ for most GRBs, then the
  total energy that is required could be lowered.  However GRBs at
  $z\ll 1$ are rare due to the smaller available volume, and a very
  large Compton parameter, $Y\gtrsim 10^3$, would still be required.}
In addition to this, $Y\sim (\epsilon_e/\epsilon_B)^{1/2}$ for $Y\gg
1$, where $\epsilon_e$ ($\epsilon_B$) is the fraction of the internal
energy behind the shock in the relativistic electrons (magnetic
field). Therefore, $Y\gtrsim 10^3$ implies $\epsilon_B\lesssim
10^{-6}\epsilon_e\sim 10^{-7}-10^{-6}$, which is an extremely low
value both compared to the values expected from the magnetic field
advected with the ejecta from the central source (Spruit, Daigne \&
Drenkhahn 2001) and compared to the magnetic field expected to be
produced at the internal shocks themselves (Medvedev \& Loeb 1999).
Together with the difficulty mentioned above in explaining the
different temporal behavior of the two spectral components, we find
that this explanation can be ruled out.

Another emission mechanism, which was favored by G03, is a hadronic
cascade, initiated by protons that are accelerated in the internal
shocks up to $\sim 10^{20}\;$eV, and make photomeson interactions with
the synchrotron photons, producing pions which decay into high energy
photons. The latter pair produce with lower energy photons creating a
cascade.  The duration of the emission from this cascade is similar to
that of the lower energy component ($T_{\rm GRB}$), since adiabatic
cooling becomes significant on the time scale of a single pulse, that
is typically $\ll T_{\rm GRB}$. Also, the spectral slope is too soft,
$\alpha\approx 1$ (Begelman, Rudak \& Sikora 1990; Peer \& Waxman
2003, in preparation). Therefore, this option does not work well.

In order to explain the longer duration of the high energy spectral
component, one can turn to models where additional interactions occur
outside of the internal shocks region, on the way to the observer,
causing a delay in the arrival time of the high energy photons. One
example for such a model features interactions of high energy photons
emitted in the internal shocks with the cosmic IR background,
producing $e^\pm$ pairs which upscatter CMB photons (Dai \& Lu 2002;
Guetta \& Granot 2003a). However, the expected duration of this
emission is $\gtrsim 10^3\;$s, and the (time integrated) spectral
slope is too soft, $F_\nu\propto\nu^{-\alpha}$ with
$\alpha=\frac{p+2}{4}\approx 1-1.25$, where $p\sim 2-3$ is the electron
power law index.  Another mechanism that produces delayed high energy
emission is the interaction of ultra-high energy cosmic rays that are
accelerated in the internal shocks with CMB photons, which produces by
cascading GeV-TeV photons (Waxman \& Coppi 1996).  However the typical
time scale for this emission is hours to days, and the spectral slope
is again too soft, $\alpha\approx 0.8$.  Therefore these two
mechanisms do not provide a good explanation for the high energy
component in GRB 941017.

\section{Reverse Shock and Early Forward Shock}
\label{sec:RS}

Since the reverse shock is a physically distinct region from the
internal shocks that emit the lower energy component in GRB 941017,
the different temporal behavior of the two components arises naturally
in this scenario.\footnote{A more detailed analysis of the high energy
  emission from the reverse shock and early forward shock is left for
  a different work (Granot \& Guetta in preparation), while here we
  briefly mention features that are relevant for GRB 941017.}  The
relevant parameters that determine the interaction of the shell of
ejecta with the ambient medium are its (isotropic equivalent) energy
$E$, initial Lorenz factor $\eta$, initial width $\Delta_0$, and the
external mass density profile, which for simplicity is assumed to be a
power law with the radius $R$, $\rho_{\rm ext}=AR^{-k}$.  The most
physically interesting external density profiles are $k=0$ and $k=2$,
that correspond to a constant density medium (like the ISM) and a
stellar wind of a massive star progenitor, respectively.  The behavior
of the system divides into two limits according to the value
of\footnote{This parameter is a power of the usual parameter
  $\xi=\tilde{\xi}^{1/(2-k)(3-k)}$ (Sari \& Piran 1995), and is more
  convenient to work with, as it is well behaved at $k=2$ which is of
  physical interest.}  $\tilde{\xi}\sim
E/Ac^2\Delta_0^{3-k}\eta^{2(4-k)}$. For $\tilde{\xi}>1$, 
or the `thick shell' case, the emission from the reverse shock peaks
after the end of the prompt GRB, and there is a temporal separation
between the two (Sari 1997). For $\tilde{\xi}<1$, or the `thick shell'
case, there is an overlap between the reverse shock emission and the
prompt GRB. As in GRB 941017 there is a significant temporal overlap
between the two spectral components, a thick shell is clearly the
relevant case here.

For a thick shell, the reverse shock is relativistic, either from the
very start for $k=2$, or from $t_N$ for $k<2$, where $t_N\approx
10\zeta^{3/2}E_{54}^{1/2} n_0^{-1/2}\eta_{2.5}^{-4}T_{80}^{-1/2}\;{\rm
  s}$ for $k=0$, where $\zeta=(1+z)/2$, $n=n_0\;{\rm cm^{-3}}$ is the
external density, and $T_{\rm GRB}= (1+z)\Delta_0/c=80T_{80}\;$s.
Unless specified otherwise, $Q_x=Q/[10^x\times({\rm the\ c.g.s\ units\ 
  of}\ Q)]$.  The reverse shock finishes crossing the shell at
$t_E\sim 2T_{\rm GRB}=160T_{80}\;$s.  After $t_E$ no new electrons
are accelerated and the hot electrons quickly cool, both adiabatically
and radiatively, so that the observed emission decays rapidly.
This provides roughly the right time scale for the high energy
component in GRB 941017.

There are two emitting regions: the shocked ejecta behind the reverse
shock, and the shocked external medium behind the forward shock. This
implies four IC components (Wang, Dai, \& Lu 2001b),
where the scattering electrons and seed synchrotron photons can be
from either of these two regions. The SSC emission 
(where both the seed photons and scattering electrons are from the
same region)
from the forward shock peaks at $\sim\;$TeV energies
and is thus not relevant here.\footnote{This component, in the thin
  shell case, was suggested by M\'esz\'aros \& Rees (1994) as a
  possible explanation for the higher energy emission ($\sim
  1-10\;$GeV) that was detected by the EGRET spark chamber in a few
  GRBs (Hurley et al.  1994) over longer time scales (up to $\sim
  1.5\;$hr for GRB 940217).  For a thick shell, this component peaks
  at very high energies ($\gamma_m^2 h\nu_m\sim\;$TeV) at $t<t_E$,
  while at $t>t_E$ it decays with time as $\gamma_m^2 h\nu_m\propto
  t^{-(18-5k)/2(4-k)}$ and $(\nu F_\nu)_{\rm max}\propto t^{-1}$,
  making it difficult to explain bright high energy emission at $t\gg
  T_{\rm GRB}$.} The external Compton (EC) processes, where the seed
photons are emitted in the reverse shock and the scattering electrons
are in the forward shock, or vice versa, have a typical photon energy
$\sim 10-100\;$GeV that implies a total energy $\sim 10^2-10^3$ times
higher than that in the observed energy range.  Nevertheless, they may
still be viable options for somewhat less typical parameters. For
concreteness, we will concentrate on the best candidate -- SSC
emission from the reverse shock, that naturally peaks at a few
hundred MeV (Wang, Dai, \& Lu 2001a), so that the total energy is
comparable to that in the observed range.

For an external density typical of the ISM ($n_0\sim 1$) the reverse
shock is in the slow cooling (SC) regime, while for a stellar wind
with $A=5\times 10^{11}A_*\;{\rm gr\;cm^{-1}}$ there is fast cooling
(FC; e.g. Chevalier \& Li 2000).  For simplicity, we neglect
synchrotron self absorption, and the effects of Klein-Nishina and
opacity to pair production, which are typically not important in the
observed energy range for GRB 941017.  The SSC spectrum in the two
cooling regimes is given to the lowest order approximation by the
following broken power law form:
\begin{equation}\label{nuFnu_SSC_SC}
\frac{\nu F^{\rm IC}_\nu({\rm SC})}{Y(\nu F_\nu)^{\rm syn}_{\rm max}} 
=\left\{\matrix{
\left(\frac{\nu^{\rm IC}_m}{\nu^{\rm IC}_c}\right)^{\frac{3-p}{2}}
\left(\frac{\nu}{\nu^{\rm IC}_m}\right)^{\frac{4}{3}} & \nu<\nu^{\rm IC}_m \cr & \cr
(\nu/\nu^{\rm IC}_c)^{(3-p)/2} & \nu^{\rm IC}_m<\nu<\nu^{\rm IC}_c \cr & \cr
(\nu/\nu^{\rm IC}_c)^{(2-p)/2} & \nu>\nu^{\rm IC}_c }\right. \ ,
\end{equation}
\begin{equation}\label{nuFnu_SSC_FC}
\frac{\nu F^{\rm IC}_\nu({\rm FC})}{Y(\nu F_\nu)^{\rm syn}_{\rm max}} 
=\left\{\matrix{
\left(\frac{\nu^{\rm IC}_c}{\nu^{\rm IC}_m}\right)^{\frac{1}{2}}
\left(\frac{\nu}{\nu^{\rm IC}_c}\right)^{\frac{4}{3}} &
\nu<\nu^{\rm IC}_c \cr & \cr
(\nu/\nu^{\rm IC}_m)^{1/2} & \nu^{\rm IC}_c<\nu<\nu^{\rm IC}_m \cr & \cr
(\nu/\nu^{\rm IC}_m)^{(2-p)/2} & \nu>\nu^{\rm IC}_m }\right. \ ,
\end{equation}
\begin{equation}\label{nuFnu_max_syn}
(\nu F_\nu)^{\rm syn}_{\rm max} =
\frac{10^{-6}f_*}{(1+Y)} \zeta a g \epsilon_{e,0.3}
E_{54} T_{80}^{-1} d_{L28}^{-2}\;\frac{{\rm erg}}{\rm cm^2\;s}\ ,
\end{equation}
where $f_*\approx 0.4-0.8$ for $0\lesssim k\lesssim 2$, $d_L$ is the
luminosity distance to the GRB, $g=3(p-2)/(p-1)$,
$a=\min[1,(\gamma_m/\gamma_c)^{p-2}]$, and
$\epsilon_{e,0.3}=\epsilon_e/0.3$. In order to minimize the total
required energy we would like $Y(\nu F_\nu^{\rm IC})^{\rm syn}_{\rm
  max}$ to be close to the maximal observed value of $\nu F_\nu$ for
the high energy component, i.e. $\sim 3\times 10^{-6}\;{\rm
  erg\;cm^{-2}\;s^{-1}}$.  Together with the roughly constant flux,
and assuming $Y\gtrsim 1$, this requires a peak frequency,
$\max(\nu^{\rm IC}_m,\nu^{\rm IC}_c)\sim\;$a few hundred MeV, and
roughly constant in time.  Since $\nu^{\rm IC}_m\propto
t^{2(1-k)/(4-k)}$ and $\nu^{\rm IC}_c\propto t^{6(k-1)/(4-k)}$, both
frequencies are constant in time for $k=1$, for which
\begin{eqnarray}
\label{nu_m_IC}
h\nu_m^{\rm IC} &=& 160\zeta^{-3/2}g^4\epsilon_{B,-2}^{1/2}\epsilon_{e,0.5}^4
E_{54}^{-1/2}T_{80}^{1/2}A_{-5}\eta_{3}^{4}\;{\rm MeV}\ ,\quad
\\ \label{nu_c_IC}
h\nu_c^{\rm IC} &=& 0.02\zeta^{-3/2}\epsilon_{B,-2}^{-3/2}
\epsilon_{e,0.5}^{-2}E_{54}^{-1/2}T_{80}^{1/2}A_{-5}^{-3}\;{\rm eV}\ .
\end{eqnarray}
However, if the lower energy component is attributed to synchrotron
emission from the internal shocks, then 
\begin{equation}\label{nu_syn_IS}
E_{\rm peak}=h\nu_m^{\rm syn} = 1.2 \zeta^{-1/2} g^2 \epsilon_{B,-2}^{1/2}
\epsilon_{e,0.5}^{2} E_{54}^{1/2} T_{80}^{-1/2}
\eta_{3}^{-2} t_{v,-3}^{-1}\;{\rm keV}\ ,
\end{equation}
where $t_v$ is the variability time of the source.  The finite size
$l_s$ of the central source implies $t_v\gtrsim 10^{-4}(l_s/\,30\,{\rm
  km})\;$s, so that $t_{v,-3}\gtrsim 0.1$ for a reasonable source
size. Since the reverse shock and the internal shocks propagate into
the same ejecta, it is reasonable to expect similar values of
$\epsilon_B$ and $\epsilon_e$. For $E=10^{54.5}\;$erg, $\eta=500$,
$t_v=10^{-4}\;$s, $A=10^{-3.5}\;{\rm gr\;cm^{-2}}$, and the other
parameters at their fiducial values, we obtain $E_{\rm peak}\sim
100\;$keV, $h\nu_m^{\rm IC}\sim 200\;$MeV and $\nu_m^{\rm IC}
F_{\nu_m^{\rm IC}}\sim 2\times 10^{-6}{\rm erg\;cm^{-2}\;s^{-1}}$.
Therefore, reasonable parameters\footnote{This value of $A$ implies
  $n\sim 200\;{\rm cm^{-3}}$ at $R\sim 10^{18}\;$cm, which is a bit
  high. However, for these parameters the reverse shock finishes
  crossing the shell at $R=2.8\times 10^{16}\;$cm, so that if
  $k\approx 1$ only at small radii $\lesssim (0.3-1)\times
  10^{17}\;$cm, and $k\approx 2$ at larger radii, then we can have
  $n\sim 6-20\;{\rm cm^{-3}}$at $R\sim 10^{18}\;$cm which is more
  reasonable. Such a variation in $k$ with radius might possibly
  result from a variation in the mass loss rate or wind velocity of
  the massive star progenitor toward the end of its life.} can yield a
reasonable fit to the data, with one major drawback: the spectral
slope is $\alpha=1/2$, which is only marginally consistent with the
observed value of $\alpha\sim 0$.  In order not to see the synchrotron
emission from the forward shock at $t\lesssim 200\;$s, we need
$Y\gtrsim 100$ in the forward shock, or\footnote{This is not so
  extreme, as the external medium typically has a weak magnetic field,
  and there is also evidence from a recent GRB (021211) suggesting
  that $\epsilon_B$ is smaller in the forward shock compared to the
  reverse shock (Kumar \& Panaitescu 2003).} $\epsilon_B\lesssim
10^{-4}$.

It is possible to bring the spectral slope closer to $\alpha\sim 0$ if
$h\nu_c^{\rm IC}\sim 1\;$MeV, since then $\alpha$ would gradually
change from $-1/3$ to $1/2$ (e.g. Sari \& Esin 2001), and could pass
for a power law with $\alpha\sim 0$ in the relatively narrow range
between a few MeV to $\lesssim 200\;$MeV.  However $h\nu_c^{\rm
  IC}\sim 1\;$MeV, requires $\epsilon_B\lesssim 10^{-4}$ and
$A\lesssim 10^{-6}\;{\rm gr\; cm^{-2}}$, which in turn require
$\eta\sim 10^{3.5}$ in order to keep $h\nu_m^{\rm IC}\sim 200\;$MeV.
This would imply a very low $E_{\rm peak}$ for the internal shocks, if
it is identified with $h\nu_m^{\rm syn}$. This problem can be solved
if for this GRB, unlike most GRBs, the prompt GRB emission is SSC
emission from the internal shocks, rather than synchrotron emission
(Panaitescu \& M\'esz\'aros 2000).  This picture works, but with
somewhat extreme parameters. For example, with $A=10^{-6.5}\;\;{\rm
  gr\; cm^{-2}}$, $\eta=10^{3.5}$, $g=1.5$ and $t_v=10^{-1.5}\;$s, we
obtain $h\nu_m^{\rm IC}\approx 250\;$MeV, $h\nu_c^{\rm IC}\approx
1\;$MeV, and $E_{\rm peak}^{\rm SSC}\approx 200\;$keV. The fact that
extreme parameters are required can explain why such a high energy
component is relatively rare in GRBs.

GRB 941017 was exceptionally bright with a fluence of $f=1.6\times
10^{-4}\;{\rm erg\; cm^{-2}}$ (Preece et al.  2000), comparable to the
famous GRB 990123 with $f=2.7\times 10^{-4}\;{\rm erg\; cm^{-2}}$,
$z=1.6$ and an isotropic equivalent energy output in $\gamma$-rays of
$1.4\times 10^{54}\;$erg (Kulkarni et al. 1999).  For a reasonable
radiative efficiency ($\sim 20\%$) and $z\sim 1$ this
implies\footnote{In this picture, GRB 941017 was likely collimated
  into a very narrow jet, just like GRB 990123, so that the true
  kinetic energy in the ejecta shell is probably $\sim 2-3$ orders of
  magnitude lower than the isotropic equivalent value, i.e.  $\lesssim
  10^{52}\;$erg, similar to or slightly lower than the value estimated
  for GRB 990123.}  $E\sim 10^{54}-10^{55}\;$erg for GRB 941017, in
agreement with the values used above. This interpretation implies a
bright prompt optical emission, similar to the `optical flash' in GRB
990123, for GRBs like 941017 with a bright high energy spectral
component.\footnote{in GRB 990123 the `optical flash' emission reached
  $\sim 1\;$Jy (or 9th mag; Akerlof et al.  1999), while for GRB
  941017 we estimate the the prompt optical emission to be $\sim
  5\;$Jy (or $\sim 7$th mag).} If the prompt GRB is due to SSC
emission from the internal shocks, then the synchrotron component
should peak near the optical and produce bright optical emission with
the same temporal behavior as the prompt
GRB. 

An alternative explanation mentioned by G03 arises in the supranova
model, where the GRB is expected to occur inside a pulsar wind bubble
(PWB; K\"onigl \& Granot 2002). The PWB photons can be upscattered by
the electrons behind the reverse and forward shocks, producing high
energy EC emission (Inoue, Guetta \& Pacini 2003; Guetta \& Granot
2003b).  However the flux level and the temporal behavior of the EC
component are not consistent with the data.

Another emission process that was considered by G03 is the hadronic
cascade.  Protons may be accelerated in the reverse shock up to $\sim
10^{20}\;$eV and can carry an energy comparable to the that in
$\gamma$-rays. Most of this energy may be converted into pions,
through photomeson interactions, if the shell is significantly
decelerated as happens for ``thick'' shells. The pions decay into high
energy photons which pair produce with lower energy photons thus
generating a cascade.  However, as in the case of the internal shocks,
the spectral slope is too soft, $\alpha\approx 1$
(Begelman, Rudak \& Sikora 1990; Peer \& Waxman 2003, in preparation).

\section{Discussion}
\label{sec:dis}

We have analyzed different possible explanations for the high energy
spectral component detected in GRB 941017, and find that it is hard to
explain. Most models fail quite badly.  The only reasonable
explanation we could find is emission from the reverse shock or
possibly from the very early forward shock. In this picture the high
energy component is emitted at a different physical region than the
lower energy component (i.e.  the prompt GRB that is emitted in the
internal shocks). This naturally explains the different temporal
behavior of the two components. The long duration of the GRB suggests
that we are in the `thick shell' case, which also accounts for the
temporal overlap between the two components and provides the right
time scale for the duration of the high energy component. Therefore,
we are relatively confident that the high energy component is emitted
from the reverse shock (or possibly from the very early forward shock,
while the reverse shock is still going on).

The most promising emission mechanism is synchrotron self-Compton
(SSC) emission from the reverse shock .  The spectral slope in this
picture is most naturally $F_\nu\propto\nu^{-\alpha}$ with
$\alpha=1/2$, which is only marginally consistent with the observed
value of $\alpha\sim 0$. This might suggest that an alternative or
additional emission mechanism is involved here. Nevertheless,
$\alpha\approx 0$ can be obtained for pure SSC emission from the
reverse shock with somewhat extreme parameters, for which the prompt
GRB is attributed to SSC emission from the internal shocks, rather
then synchrotron emission which is usually responsible for the prompt
GRB. This might explain why such a high energy component appears to be
rare among GRBs.

In this picture GRBs with a similarly bright high energy component
should be accompanied by a bright optical flash, as bright or even
brighter than in GRB 990123. The fact that most GRBs are not
accompanied by optical flashes of such brightness (Kehoe et al. 2001)
is nicely consistent with such a bright high energy component being
similarly rare.\footnote{A bright optical flash from the reverse shock
  should generally be accompanied by a bright high energy component.
  However, the values of the peak flux and peak energy can vary
  considerably between different GRBs, and are not related to the
  optical emission in a very simple way. Therefore, the lack of
  detection of a similar high energy component in GRB 990123 (e.g.
  Briggs et al. 1999) is perfectly consistent with this picture.}

Future missions such as GLAST,\footnote{see
  http://glast.gsfc.nasa.gov/.} will have a better sensitivity and a
wider energy range (up to $300\;$GeV for GLAST) and
should provide a much clearer picture as to how common such high
energy spectral components are in GRBs. The wider energy range may
cover the peak of $\nu F_\nu$, and thus tell us how much energy is in
the high energy component. A more accurate measurement of the spectrum
and the temporal behavior would help constrain the different models
and pinpoint the source of the high energy emission.

\acknowledgements

We thank Tsvi Piran, Eli Waxman and Asaf Peer for useful discussions.  JG is
supported by the W.M. Keck foundation, and by NSF grant PHY-0070928.
DG acknowledges the RTN ``Gamma-Ray Bursts: An Enigma and a Tool'' for 
supporting this work.

\end{document}